\documentclass[12pt,a4paper,final]{iopart}
\usepackage{iopams}
\usepackage{bm}
\usepackage{caption}
\expandafter\let\csname equation*\endcsname\relax
\expandafter\let\csname endequation*\endcsname\relax

\usepackage{amsmath}
\usepackage{amssymb}
\usepackage{amsthm}
\usepackage[utf8]{inputenc}
\usepackage{graphicx}         
\usepackage{bm}               
\usepackage{physics}
\usepackage{color}
\usepackage{xcolor}
\usepackage{soul}
\usepackage{cite}
\usepackage{dsfont}
\usepackage{subfig}

\usepackage[breaklinks=true,colorlinks=true,linkcolor=blue,urlcolor=blue,citecolor=blue]{hyperref}

\renewcommand{\eqref}[1]{Eq. (\ref{#1})}
\newcommand{\Cov}{\mathrm{Cov}}

\begin{document}
\newcommand{\nl}{{\cal N}}

\title{Quantum Expectation Identities for the Three-State Model of a Molecular Domain}

\vskip 5mm

\noindent {\bf Boris Maul\'en }$^{*,1,2}$, {\bf
Roberto C. Bochicchio}$^{3,4}$

\vskip 5mm

{\small

\noindent $^{1}$ {\it Departamento de F\'isica, FCFM, Universidad de Chile, Santiago, 8370448, Chile}

\noindent $^{2}$ {\it Departamento de F\'isica y Astronom\'ia, Facultad de Ciencias Exactas, Universidad Andres Bello, Sazi\'e 2212, piso 7, 8370136, Santiago, Chile}

\noindent $^{3}$ {\it Universidad de Buenos Aires, Facultad
de Ciencias Exactas y Naturales, Departamento de F\'{\i}sica, Ciudad
Universitaria, 1428, Buenos Aires, Argentina}

\noindent $^{4}$ CONICET - {\it Universidad de Buenos Aires,
Instituto de F\'{\i}sica de Buenos Aires (IFIBA) Ciudad
Universitaria, 1428, Buenos Aires, Argentina}

}

\vskip 5mm

\noindent {\bf Abstract}: The electronic distribution of a molecular domain is examined in this study. A theoretical formulation of quantum molecular properties is presented using the Quantum Expectation Identity theorem (QEI), with a focus on the three-state model of the density matrix for the quantum state of a molecular domain as an open system. The report examines the relationship between {\it ab initio} statistical fluctuation-correlation theorems for quantum observables  and their derivatives. We focus on three main quantities of a domain: the electronic population, its chemical potential, and its maximum capacity for accepting or donating charge with the neighbors. The analytical expressions for the quantities are presented and discussed in detail. At the end, we explore the concept of \textit{quantum purity} and its proper application in the molecular domain.

\vskip 30mm

{\small \noindent
\rule{60mm}{0.4mm} \\
\noindent $^{*}$ boris.maulen@ing.uchile.cl}


%
%
%
\newpage

\setcounter{footnote}{0}
\section{Introduction}

From a quantum perspective, a whole molecular system can be viewed as a collection of points within configuration space, encompassing both electrons and nuclei interacting primarily by Coulomb electrostatic potentials. The type of interaction for these systems has an important experimental feature: their energies and, consequently, their electronic densities behave as convex functions of the integer number of particles $M$ when it is considered as a closed system, that is to say, a neutral molecule or any of its ions \cite{PPLB,Parr_Yang_book,Geerlings, Boch_Rial_JCP_2012}.

The convexity allows for the extension of the number of particles to be treated as a continuous variable between two consecutive integer numbers of particles, $M$ and $M \pm 1$. This approach enables the investigation of variations in these quantities between these integer values. The primary conclusion from this property is that the ground state Density Matrix (DM) $\hat{\rho}$ can be rigorously represented by a two-state model within a two-dimensional Fock space. This model includes only two states: the quantum state of the neutral species {\bf X}$^{0}$ and one of the ionic species, either {\bf X}$^{+}$ or {\bf X}$^{-}$. The choice between these ionic species depends on whether their population is greater or lower than that of the neutral species, respectively \cite{PPLB,Boch_Rial_JCP_2012}.

An alternative approach to the comprehension of molecular structure, consist in considering the molecular framework as a collection of non-overlapping physical regions known as \textit{molecular domains} or basins $\Omega$ based on specific chemical criteria. In this context, two main useful tools that are able to define the molecular domains as atoms and/or chemical bonds are the so-called \textit{Quantum Theory of Atoms in Molecules} (QTAIM) \cite{Bader_book,Popelier_book}, and the \textit{Electron Localization Function} (ELF) \cite{elf1,elf2,elf3,elf4}. These models, which share a common topological framework, have no semi-empirical roots and enable us to rigorously decompose the chemical structure.

In particular, the physical domains stated in the framework of QTAIM may be represented by an atom, a functional group, or simply a moiety that interacts with other domains in the molecule, facilitating the exchange of charge (electrons) between them \cite{Bader_book,Popelier_book}. As a result, these domains can accommodate a non-integer number of electrons (fractional electronic populations): it is in this sense that the number of electrons, denoted by $\nl$, may be treated as a continuous variable. According to the above, a molecular domain is considered an open quantum system, where the electronic population $\mathcal{N}$ is defined by the relation $\mathcal{N}=N+\nu$ \cite{Wasserman,Pendas1,Boch_TCA2015,Pendas2}; $N$ represents an integer number indicating the electronic population of a molecular domain in its neutral form, while $\nu$ is the fractional charge that is either transferred to or from the domain due to its interactions with the surrounding domains \cite{bochicchio_maulen}. 

In contrast to systems with a fixed integer number of particles $N$ (the whole molecular structure), a molecular domain $\Omega$ lacks the previously mentioned convexity property. Due to the above, the density matrix $\hat{\rho}$ of $\Omega$ must be expanded using more than only two pure states\cite{Blum_book}. Thus, to address this type of system, we proposed a model for $\Omega$ in which its density matrix is expanded in terms of three states: the neutral state {\bf X}$^{0}$, and both ionic species {\bf X}$^{+q}$ and {\bf X}$^{-q}$, where $q\in \mathbb{Z^{+}}$ stands for the maximum capacity to accept or donate electrons featuring each molecular domain. So, the fraction of charge $\nu$ is restricted to the interval $-q \leq \nu \leq q$ \cite{bochicchio_maulen}. Additionally, these states serve as the basis set for the Fock space \cite{bochicchio_maulen, Boch_elec_distribution_domains}.

Chemical properties and quantum descriptors of electron distributions in molecular quantum systems are often expressed as expectation values of certain Hermitian operators. In this sense, the knowledge about the quantum state represented by the density matrix of the system is essential \cite{Coleman_Yukalov_book}. Thus, in order to make calculations on the molecular realm, the physical information in $\hat{\rho}$ can be handled by the use of some important theorems of quantum mechanics and molecular physics. In particular, the \textit{Quantum Expectation identity} (QEI) recently developed is a natural tool that allows obtaining expectation identities for quantum systems described by a mixed density matrix consistent with standard results of quantum mechanics and molecular physics, such as the Ehrenfest and Hellmann-Feynman theorems and the equilibrium fluctuation-dissipation theorem, among others \cite{maulen_2025}. In the case of molecular domains, the QEI is an adequate tool to describe the charge transfer and its derivatives, as well as the statistical correlations between certain chemical properties.

The objective of the present report is to revise the perspective of general identities related to various averages and the chemical descriptors for a molecular domain treated as a three-state quantum system, by means of the QEI and its subsequent identities.

The article is organized as follows. The second section introduces a theoretical framework for the results presented in this report, specifically focusing on the QEI and the model for describing the quantum state of molecular domains. The third section reviews the results of specific cases, examining the identities that arise using the identity operator, the  particle number operator, and the density matrix itself as test operators in the QEI. In particular, we explore the concept of quantum purity applied to a molecular domain and its dependence on the chemical potential. Finally, the concluding section summarizes the findings and presents the main observations.

\textit{A remark about the notation}: in this work, we use the symbol $\ln$ for the natural logarithm of a scalar function, and the symbol $\log$ for the natural logarithm of an operator (see Eq. (\ref{QEI})). Also, the circumflex accent or "hat" over a symbol denotes an operator, as is customary in standard literature of quantum chemistry and molecular physics.

\section{Theoretical background: Quantum expectation identity and molecular domain quantum states}
\subsection{Quantum expectation identity}

To compute expectation identities in a quantum system, such as a molecular domain, it is essential to know what the quantum state  describing the system is. For a molecular domain considered as an open quantum system having a fractional number of electrons $\mathcal{N}$, the complete description of its quantum state is given by a convex mixture of pure states, each of them associated with its corresponding integer number of electrons $N$ \cite{Blum_book,greiner}.

In previous reports, we have shown that it is possible to expand the density matrix for a molecular domain using only three pure states: the neutral pure state, representing the molecular domain in its neutral form having $N$ electrons; the cationic pure state, representing the molecular domain having ($N-q$) electrons; and the anionic pure state, representing the molecular domain having ($N+q$) electrons \cite{bochicchio_maulen,Boch_elec_distribution_domains}. Within the thermodynamical approach of Density Functional Theory, the above model for studying a molecular domain is known as the \textit{N-centered ensemble} \cite{N_center_ensemble}. Our present model, however, is essentially a canonical-like ensemble, in which the inverse temperature is replaced by the chemical potential and the Hamiltonian operator is replaced by the number operator. Since our approach is based solely on quantum mechanics, the effects of temperature are outside the scope of this work.

In order to derive relations between some observables in a molecular domain, we use the QEI theorem. It establishes a relation between the derivative of the expectation value of some observable with other expectation vale taken in the same state of knowledge given by $\hat{\rho}$ \cite{maulen_2025}. Namely,

\begin{equation}
\dfrac{\partial}{\partial \lambda} \left\langle \hat{A}(\lambda)
\right\rangle_{\hat{\rho}} = \left\langle \dfrac{\partial \hat{A}(\lambda) }
{\partial \lambda} \right\rangle_{\hat{\rho}}+\left\langle \hat{A}(\lambda)\dfrac{\partial \log \hat{\rho}(\lambda) }{\partial \lambda} \right\rangle_{\hat{\rho}}. \label{QEI}
\end{equation}
In Eq. (\ref{QEI}), $\hat{A(\lambda)}$ is a quantum observable, called the \textit{test operator}, which may or may not depend on a continuous parameter $\lambda$, and $\hat{\rho}(\lambda)$ is the density matrix representing the accessible quantum states of the system. The expression $\langle  ... \rangle_{\hat{\rho}}$ means the quantum average over the state density matrix $\hat{\rho}$. In particular, the parameter $\lambda$, also called the \textit{derivation parameter}, may be a parameter contained in the observables (e.g. charge, mass, time, etc.) or may be parameters that accompany the observables as the Lagrange multipliers associated with different constraints in a maximization procedure of the von Neumann entropy (e.g. chemical potential, inverse temperature, etc.). An essential requirement for the observable $\hat{A}$ to make the above equality hold is that this must be compatible with the density matrix of the system, i.e.

\begin{equation}
\left[ \hat{A},\hat{\rho}\right]=0.
\end{equation}
In addition, the density matrix $\hat{\rho}$ must be a non-singular operator, which is achieved by considering its non-zero eigenvalues only: in a molecular domain spanned by only three pure states, the above is fulfilled. See Ref. \cite{maulen_2025} for more details.

\subsection{Statistical three-state model for a molecular domain}
Within the inferential approach of E. T. Jaynes of probabilities, the least-biased statistical distribution of the states in a quantum system is obtained by maximizing the von-Neumann entropy, under certain constraints \cite{jaynes1,jaynes2,maxent_rdm}. For a molecular domain treated as an open quantum system exchanging particles (electrons) with its surroundings, a natural constraint, in addition to the trivial constraint (normalization), is the fluctuation of the particle number $N$ around its mean $\mathcal{N}$. In this maximization procedure, each constraint is linked with a specific Lagrange multiplier. In particular, for the constraint of the fluctuation of the particle number, the electronic chemical potential $\gamma$ is an appropriate Lagrange multiplier. The parameter $\gamma$ should not be confused with the chemical potential $\mu$ of \textit{Conceptual Density Functional Theory} (CDFT) \cite{Parr_Yang_book,Geerlings}. The former, $\gamma$, acts as a control parameter that enables us to investigate the response of the molecular domain within an electronic environment, and may be interpreted as the chemical potential of an open system \cite{Boch_elec_distribution_domains}. In contrast, the latter, $\mu$, is always a negative quantity describing a stable electron system, as the whole molecular (closed) system.

After maximizing the entropy under the aforementioned constrains only, is obtained, as the quantum state describing the system, a density matrix of the form 
\begin{equation}
\hat{\rho}(\gamma;q)=\dfrac{e^{-\gamma \hat{M}}}{\Xi(\gamma;q)}, \label{D}
\end{equation}
where $\hat{M}$ is the number operator and $\Xi(\gamma;q)$ is the partition function,
\begin{equation}
\Xi(\gamma;q)=\Tr\left\lbrace e^{-\gamma \hat{M}} \right\rbrace.
\end{equation}
We have called this model the $\nl$-canonical quantum statistical model \cite{Boch_elec_distribution_domains} (in short $\nl$-canonical model).

The molecular domain under study may be handled as a three-state quantum system, as stated in Ref. \cite{bochicchio_maulen,Boch_elec_distribution_domains}. In this sense, once a basis of pure states for the molecular domain is given, we can use it in order to build the spectral decomposition of the density matrix (\ref{D}). In particular, we use a three-dimensional basis given by $\left\lbrace \ket{\Phi^M} \bra{\Phi^M} \right\rbrace$, with $M=N$ representing the neutral pure state, $M=N-q$ representing the cationic pure state, and with $M=N+q$ representing the anionic pure state. Thus, the spectral decomposition of the density matrix (\ref{D}) is written as
\begin{equation}
\hat{\rho}(\gamma;q)=\sum_{M=N,N\pm q} \omega^M(\gamma;q) \ket{\Phi^M}\bra{\Phi^M},
\end{equation}
where the statistical weights $\omega^M(\gamma;q)$ (eigenvalues of $\hat{\rho}_g$) are given by
\begin{equation}
\omega^M (\gamma;q)=\dfrac{e^{-\gamma M}}{\Xi(\gamma;q)},
\end{equation}
with $M=N,N\pm q$. In this three-state basis, the partition function $\Xi(\gamma;q)$, which we called the $\nl$-canonical partition function from now on, can be expressed explicitly as \cite{Boch_elec_distribution_domains}
\begin{equation}
\Xi(\gamma;q)=e^{-\gamma N} \left[  1+2\cosh\left( \gamma q \right)  \right]. \label{analytical_grand-c_partition}
\end{equation}

\section{Results and discussion}

The expectation identities from the QEI for a molecular domain using the $\nl$-canonical density matrix $\hat{\rho}(\gamma;q)$
in Eq. (\ref{QEI}) reads
\begin{equation}
\dfrac{\partial}{\partial \lambda}\left\langle \hat{A} \right\rangle_{\hat{\rho}} =\left\langle \dfrac{\partial \hat{A}}{\partial \lambda} \right\rangle_{\hat{\rho}}-\left\langle \hat{A}\hat{M} \right\rangle_{\hat{\rho}}-\dfrac{\partial \ln \Xi(\gamma;q)}{\partial \lambda} \left\langle \hat{A} \right\rangle_{\hat{\rho}}.  \label{QEI_grand_canonical}
\end{equation}
In the following discussion, we will use the chemical potential $\gamma$ found in $\hat{\rho}(\gamma;q)$ as the derivation parameter $\lambda$. We will examine the expectation identities of interest that arise from Equation (\ref{QEI_grand_canonical}) with $\lambda=\gamma$  for the following test operators:  i. $\hat{A}=\mathds{1}$, ii. $\hat{A}=\hat{M}$. Additionally, we will consider the case where $\hat{A}=\hat{\rho}$, separately due to its significance.



Regarding the convergence aspects of this ensemble, it is important to note that the density matrix presented in Eq. (\ref{D}) has a structure similar to that of a standard canonical density matrix. In this case, the Hamiltonian operator is replaced by the number operator, and the inverse temperature $\beta$ is substituted with the parameter $\gamma$. However, while the canonical distribution only converges for $\beta >  0$, our $\nl$-canonical ensemble distribution, as given by Eq. (\ref{D}), converges for both positive and negative values of $\gamma$. It is due to a molecular domain is characterized by its maximum capacity to accept or donate electrons, as stated in the introduction of this work, being this quantity ($q$) a finite number. Hence, the electron number in a domain, unlike the energy in a typical quantum system, is a completely bounded quantity. Thus, in the limit $\gamma \rightarrow \mp \infty $, the expectation $\left\langle \hat{M} \right\rangle$ reaches its maximum/minimum value given by $N \pm q$, respectively. At this point, it is essential to remark that the statistical model described by the density matrix in Eq. (\ref{D}) does not incorporate temperature, as we are examining a purely quantum scenario. As a result, the expectation identities derived in this context are structurally similar to those found in standard statistical mechanics, with the absence of the inverse temperature \cite{kubo_book}.

\subsection{Identities for the particle number}

For the case $\hat{A}=\mathds{1}$ in Eq. (\ref{QEI_grand_canonical}), we have that
\begin{equation}
\dfrac{\partial}{\partial \gamma} \left\langle \mathds{1} \right\rangle_{\hat{\rho}}=0\hspace{0.4cm}\text{and}\hspace{0.4cm}\left\langle \dfrac{\partial \mathds{1}}{\partial \gamma} \right\rangle_{\hat{\rho}}=0.
\end{equation}
Then,
\begin{equation}
\left\langle \hat{M} \right\rangle_{\hat{\rho}}=-\dfrac{\partial \ln \Xi(\gamma;q)}{\partial \gamma}, \label{mean_value_M}
\end{equation}
where {\small{$\left\langle \hat{M}
\right\rangle_{\hat{\rho}}=\mathcal{N}$}}. The identity shown in Eq. (\ref{mean_value_M}) is the standard way to obtain the mean particle number from the canonical partition function. The importance of this identity lies in the fact that it represents a
working expression for computing the mean electron number of a molecular domain modeled by the $\nl$-canonical density matrix Eq. (\ref{D}).

Using the analytic expression for the $\nl$-canonical partition function for the three-state basis (Eq. (\ref{analytical_grand-c_partition})) in the identity (\ref{mean_value_M}), we find that the mean electron number within this statistical model turns out to be
\begin{equation}
\left\langle \hat{M} \right\rangle_{\hat{\rho}}=N- \dfrac{2q \sinh(\gamma q)}{2\cosh(\gamma q)+1}, \label{mean_value_M_2}
\end{equation}
where the second term can be identified with the charge fraction $\nu$ transferred to or from the molecular domain \cite{Boch_elec_distribution_domains}, i.e.
\begin{equation}
\nu=-\dfrac{2q \sinh(\gamma q)}{2\cosh(\gamma q)+1}. \label{nu}
\end{equation}
With Eq. (\ref{nu}) the mean electron number of a molecular domain is written simply as
\begin{equation}
\left\langle \hat{M} \right\rangle_{\hat{\rho}}=N+\nu.
\end{equation}

\begin{figure}[htp]
    \centering
    \subfloat[]{ \label{fig1:a}
        \includegraphics[width=0.47\textwidth]{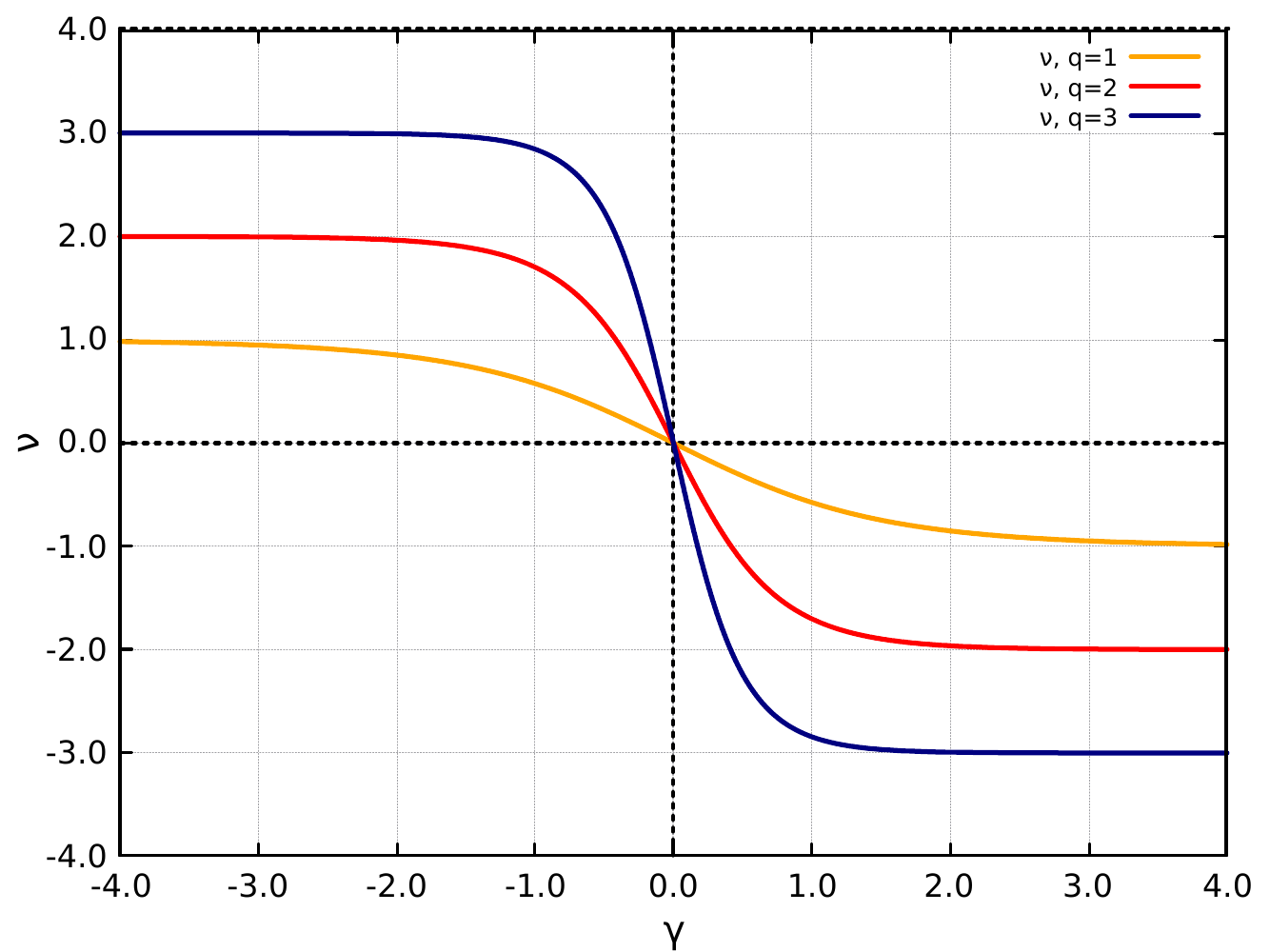}

        }
    \hfill
    \subfloat[]{  \label{fig1:b}
        \includegraphics[width=0.47\textwidth]{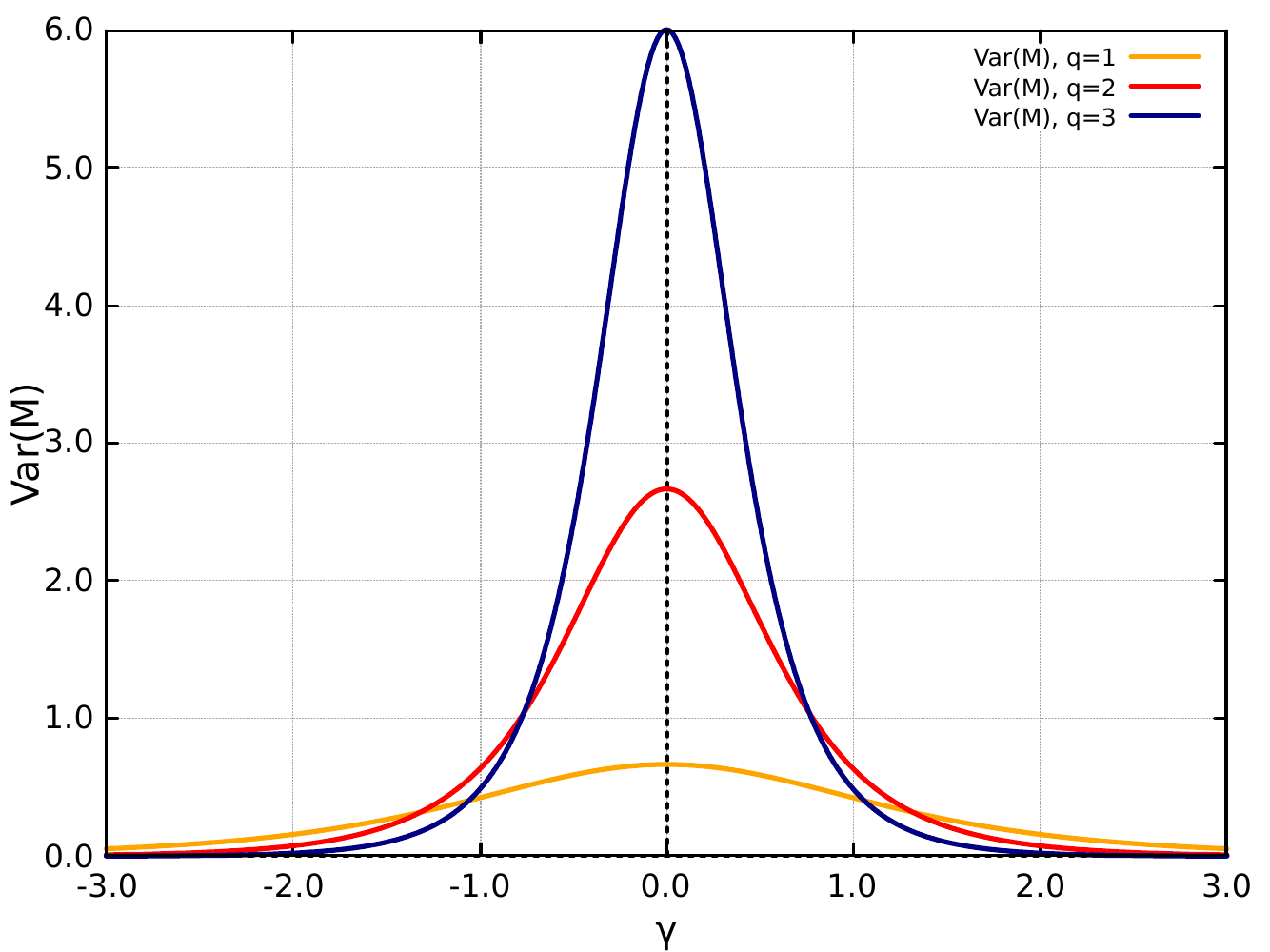}

        }
    \caption{(a) Transferred fraction of
charge $\nu$ and (b) variance (electron fluctuation) as a function of the control parameter $\gamma$,
    for different values of the maximum charge capacity $q$.} \label{fig1}
\end{figure}

In Fig. \ref{fig1:a}, by means of Eq. (\ref{nu}), we show the variation in the transferred fraction of charge $\nu$ with respect to the control parameter $\gamma$ for various maximum capacity values $q$. Moving $\gamma$ from zero to negative values, we observe that $\nu$ increases until it reaches asymptotic values representing equilibrium acceptor states (limit $\gamma \rightarrow -\infty$). On the other hand, moving from negative to positive values of $\gamma$, the charge of fraction $\nu$ decreases from its maximum $q$ to its minimum $-q$ values: $q$ and $-q$ represent the asymptotic limits associated with acceptor/donor equilibrium states, as $\gamma$ approaches to $\mp \infty$, respectively. In this sense, the asymptotic fraction of the population transferred, $\nu$, is determined by the maximum capacity of acceptors or donors, $\pm q$. The aforementioned tendencies of $\gamma$ are in full agreement with the physical meaning of the chemical potential in the grand canonical model of statistical mechanics. In such a context $\mu$ is related to the free energy via \cite{greiner,mu_physrep_kaplan}
\begin{equation}
\mu=\left(\dfrac{\partial F}{\partial \nl}\right)_{T,V}.
\end{equation}
We observe, from the above relation, that a positive chemical potential entails an increment in the free energy with the number of particles $\nl$, thus favoring the release of electrons from the domain, whereas a negative chemical potential entails a decrease in the free energy with $\nl$, thus favoring the entry of electrons to the domain.

For the choice $\hat{A}=\hat{M}$ in Eq. (\ref{QEI_grand_canonical}), we have
\begin{equation}
\left\langle \dfrac{\partial \hat{M} }{\partial \gamma} \right\rangle_{\hat{\rho}}=0,
\end{equation}
and using Eq. (\ref{mean_value_M}), Eq. (\ref{QEI_grand_canonical}) becomes
\begin{equation}
\dfrac{\partial}{\partial \gamma}\left\langle \hat{M} \right\rangle_{\hat{\rho}} = - \mathrm{Var}_{\hat{\rho}}\left( \hat{M} \right), \label{FDT_grand_canonical}
\end{equation}
where 
\begin{equation}
\mathrm{Var}_{\hat{\rho}}\left( \hat{M} \right) \equiv \left\langle \hat{M}^2 \right\rangle_{\hat{\rho}}-\left\langle \hat{M} \right\rangle^2_{\hat{\rho}}
\end{equation}
is the variance in the particle number.
Eq. (\ref{FDT_grand_canonical}) may be understood within the broader context of many-body physics. Specifically, the variation of the mean particle number with the control parameter $\gamma$ can be viewed as an \textit{electronic susceptibility}, denoted by $\chi(\gamma)$ \cite{Fetter_Walecka}. Within this context, Eq. (\ref{FDT_grand_canonical}) can be interpreted as an instance of the fluctuation-dissipation theorem applied to the particle number in a molecular domain in equilibrium with its surroundings. The fluctuation in the number of electrons, represented by the variance in $\hat{M}$, arises from the interaction between the domain and its environment. This fluctuation is linked to the dissipation driven by a changes in the chemical potential $\gamma$ of the domain.



\noindent

In addition, Eq. (\ref{FDT_grand_canonical}) enables us to determine the variance in the particle number, once we have established the mean particle number. In this context, by taking the derivative of the mean particle number, as presented in equation (\ref{mean_value_M_2}), with respect to the chemical potential $\gamma$, is obtained
\begin{equation}
\mathrm{Var}_{\hat{\rho}}\left( \hat{M}\right) =2q^2 \dfrac{\cosh(\gamma q)+2}{\left(2\cosh(\gamma q)+1 \right)^2}. \label{variance_final}
\end{equation}

Fig. \ref{fig1:b} displays the variance (dispersion) $\mathrm{Var}_{\hat{\rho}}(\hat{M})$ of the number of particles $\hat{M}$ in the state $\hat{\rho}$ of the domain, for different $q$ values, as a function of the chemical potential $\gamma$ (Eq. (\ref{variance_final})). The variance reaches its maximum value at $\gamma=0$, which indicates that the electrons neither have a tendency to enter nor to leave the domain. There is no cost in energy to accept or to donate an electron, whence a maximum fluctuation is observed. On the other hand, the limits of $\mathrm{Var}_{\hat{\rho}}$ as $\gamma$ approaches $\mp \infty$ converge to $0$, corresponding to the pure states of $N \pm q$, respectively: the variance vanishes for both acceptor and donor regimes. Note that these states are not physical and cannot be reached. Additionally, Fig. \ref{fig1:b} shows that the dispersion increases with $q$, indicating that it becomes greater as the maximum amount of charge that can be donated or accepted in the domain increases.

\subsection{Identities for the \textit{quantum purity}}
Let us now discuss the fundamental case previously mentioned: $\hat{A} = \hat{\rho}$. As explained above, the quantum state of a molecular domain can be described by a mixed state that consists of a neutral state (with $N$ electrons) and two ionic states (with $N \pm q$ electrons). Consequently, the density matrix is represented by three single projectors \cite{bochicchio_maulen,Boch_elec_distribution_domains}, i.e.

\begin{equation}
\hat{\rho}(\gamma;q)=\omega^{N-q}\ket{\Phi^{N-q}}\bra{\Phi^{N-q}}+\omega^{N}\ket{\Phi^{N}}\bra{\Phi^{N}}+\omega^{N+q}\ket{\Phi^{N+q}}\bra{\Phi^{N+q}}.
\end{equation}

A method to quantify the purity of an accessible molecular state is introduced through the concept of \textit{quantum purity} $\varphi$, defined as a real quantity according to \cite{wilde,trace_algebra}

\begin{equation}
\varphi \equiv \Tr\left\lbrace \hat{\rho}^2 \right\rbrace. \label{purity}
\end{equation}
For a pure state, the density matrix is idempotent, meaning that $\hat{\rho}^2 = \hat{\rho}$, thus $\varphi = 1$. In the opposite way, for a homogeneous mixture of $n$ different pure states, $\varphi$ reaches its minimum value of $\varphi = 1/n$, representing the less pure case \cite{purity_pennini,trace_algebra}.

An interesting application of the QEI in the context of quantum molecular domains arises when using the density matrix of the system $\hat{\rho}$ as the test operator $\hat{A}$. In such a case, Eq. (\ref{QEI_grand_canonical}) reads
\begin{equation}
\dfrac{\partial}{\partial \gamma} \left\langle \hat{\rho} \right\rangle_{\hat{\rho}} = \left\langle \dfrac{\partial \hat{\rho} }{\partial \gamma} \right\rangle_{\hat{\rho}}+\left\langle \hat{\rho}\dfrac{\partial \log \hat{\rho}}{\partial \gamma} \right\rangle_{\hat{\rho}},
\end{equation}
where the expectation of the density matrix is nothing but the purity associated to the corresponding quantum state defined above
\begin{equation}
\left\langle \hat{\rho} \right\rangle_{\hat{\rho}}=\Tr\left\lbrace \hat{\rho}\hat{\rho} \right\rbrace =\varphi.
\end{equation}
In Ref. \cite{maulen_2025} has been shown that
\begin{equation}
\left\langle \dfrac{\partial \hat{\rho} }{\partial \gamma} \right\rangle_{\hat{\rho}}=\left\langle \hat{\rho}\dfrac{\partial \log \hat{\rho}}{\partial \gamma} \right\rangle_{\hat{\rho}}.
\end{equation}
Thus, the QEI for $\hat{A}=\hat{\rho}$ and $\lambda=\gamma$ can be conveniently rewritten as
\begin{equation}
\dfrac{\partial \varphi (\gamma;q)}{\partial \gamma} =2 \left\langle \hat{\rho}(\gamma;q)\dfrac{\partial \log \hat{\rho}(\gamma;q)}{\partial \gamma} \right\rangle_{\hat{\rho}(\gamma;q)}.
\end{equation}
A more explicit identity can be expressed using the  $\nl$-canonical density matrix (Eq. (\ref{D})) in its analytical form, within the logarithm in the last equation, which results
\begin{equation}
\begin{split}
\dfrac{\partial \varphi(\gamma)}{\partial \gamma} & =  2\left\langle \hat{\rho} \dfrac{\partial}{\partial \gamma} \left[-\gamma \hat{M}-\ln\Xi(\gamma;q)\mathds{1} \right]   \right\rangle_{\hat{\rho}}\\
& = -2 \left[ \left\langle \hat{\rho}\hat{M} \right\rangle_{\hat{\rho}} + \dfrac{\partial \ln\Xi(\gamma;q) }{\partial \gamma} \left\langle \hat{\rho} \right\rangle_{\hat{\rho}}   \right].
\end{split}
\end{equation}
Also, when using the identity about the mean particle number Eq. (\ref{mean_value_M}) in the second line of the above equation, we finally get
\begin{equation}
\dfrac{\partial \varphi(\gamma)}{\partial \gamma} = -2\hspace{0.08cm}\Cov_{\hat{\rho}}\left( \hat{\rho}, \hat{M} \right), \label{purity_FDT}
\end{equation}
where $\Cov_{\hat{\rho}}\left( \hat{\rho}, \hat{M} \right)$ is the covariance between the density matrix and the operator number of particles, which are defined by
\begin{equation}
\Cov_{\hat{\rho}}\left( \hat{\rho}, \hat{M} \right) \equiv \left\langle \hat{\rho}\hat{M} \right\rangle_{\hat{\rho}}- \left\langle \hat{\rho} \right\rangle_{\hat{\rho}}\left\langle \hat{M} \right\rangle_{\hat{\rho}},
\end{equation}
and measures the statistical correlation between both operators \cite{stat_theory}. In particular, for a pure state, we have $\varphi=1$, which leads to
$\Cov_{\hat{\rho}}\left( \hat{\rho}, \hat{M} \right)=0$.  This implies that in the specific case when the quantum system is represented by a single projector (such as a neutral state or a cationic/anionic state), the density matrix and the number operator are uncorrelated observables.

\begin{figure}[htp]
    \centering
    \subfloat[]{ \label{fig2:a}
        \includegraphics[width=0.47\textwidth]{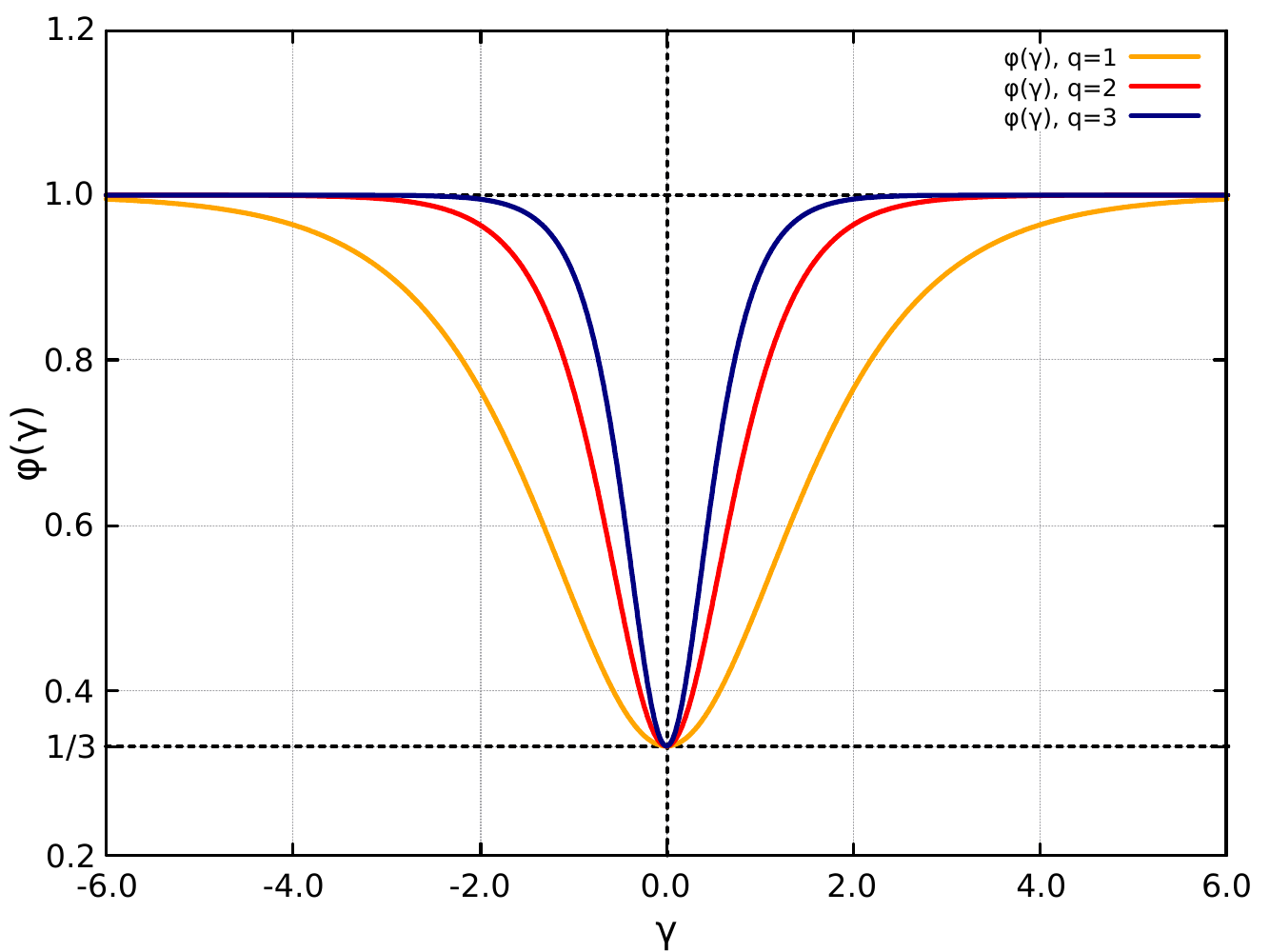}

        }
    \hfill
    \subfloat[]{  \label{fig2:b}
        \includegraphics[width=0.47\textwidth]{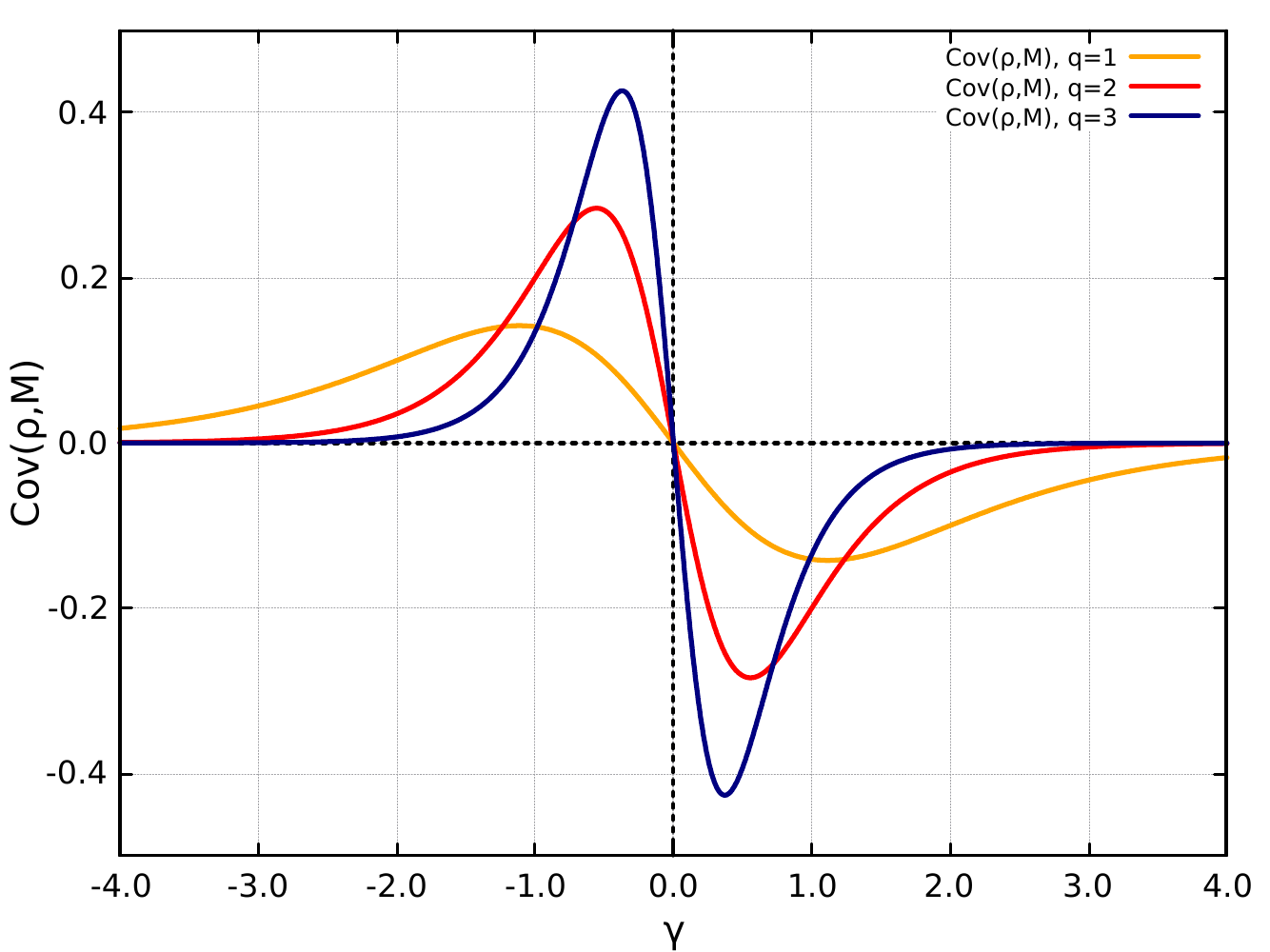}

        }
    \caption{(a) Quantum purity and (b) covariances for a molecular domain handled as a function of $\gamma$
   for diffrent values of charge $q$ ($q=1,2,3$). The covariances were obtained by use of the instance of the p-FDT shown in Eq. (\ref{purity_FDT}).} \label{fig2}
\end{figure}

The expectation identity shown in Eq. (\ref{purity_FDT}) is an instance of the \textit{purity-fluctuation-dissipation theorem} (p-FDT) stated in Ref. \cite{maulen_2025} applied to a molecular domain described by the $\nl$-canonical density matrix. In particular, Eq. (\ref{purity_FDT}) reveals that the mutual fluctuations between the quantum state, represented by $\hat{\rho}$, and the particle number is determined by the change in the quantum purity driven by a change in the electronic chemical potential $\gamma$.

To compute the quantum purity for a molecular domain explicitly using our approach, we use the analytical expression of $\hat{\rho}$ (\ref{D}) in $\varphi$. Namely 

\begin{equation}
\varphi(\gamma)=\Tr\left\lbrace  \hat{\rho}^2 \right\rbrace =  \Tr\left\lbrace \dfrac{\left( e^{-\gamma\hat{M}}\right)^2 }{\Xi^2(\gamma;q)} \right\rbrace = \dfrac{\Tr \left\lbrace \left( e^{-\gamma\hat{M}}  \right)^2  \right\rbrace }{\left( \Tr\left\lbrace e^{-\gamma\hat{M}} \right\rbrace  \right)^2 }.
\end{equation}
The numerator in the above expression is computed by expanding the trace operator over the three states that describes the molecular domain,
\begin{equation}
\Tr \left\lbrace \left( e^{-\gamma\hat{M}}  \right)^2  \right\rbrace=\sum_{M=N,N\pm q} \left(  e^{-\gamma M}  \right)^2=e^{-2\gamma N} \left[ 1+2\cosh\left( 2\gamma q\right)   \right],
\end{equation}
while the denominator is simply the $\nl$-canonical partition function squared (Eq. (\ref{analytical_grand-c_partition}))
\begin{equation}
\Xi^2(\gamma;q)=e^{-2\gamma N} \left[  1+2\cosh\left( \gamma q \right)  \right]^2.
\end{equation}
After a little algebra and using some properties of hyperbolic functions, the purity of the quantum state $\hat{\rho}$ turns out to be
\begin{equation}
\varphi(\gamma)=\dfrac{2\cosh(\gamma q)-1}{2\cosh(\gamma q)+1}. \label{purity_three_states}
\end{equation}
The covariances between $\hat{\rho}$ and $\hat{M}$ are determined by the change in the purity with the electronic chemical potential, as stated above (see Eq. (\ref{purity_FDT})). In this sense, by taking the derivative of (\ref{purity_three_states}) with respect to $\gamma$, we obtain for the covariances, the following function
\begin{equation}
\Cov_{\hat{\rho}}\left( \hat{\rho}, \hat{M} \right)=-2q \dfrac{\sinh(\gamma q)}{\left( 2\cosh(\gamma q)+1 \right)^2}. \label{cov_final}
\end{equation}


Fig. \ref{fig2:a} illustrate the degree of purity $\varphi$ of the state $\hat{\rho}$ (Eq.(\ref{purity})) as a function of $\gamma$, for some values of $q$. As $\gamma \rightarrow \pm \infty$, $\varphi \rightarrow 1$, indicating that the state approaches one of the ionic pure states: $\ket{\Phi^{N+q}}\bra{\Phi^{N+q}}$ state for $\gamma \rightarrow -\infty$ and $\ket{\Phi^{N-q}}\bra{\Phi^{N-q}}$ state for $\gamma \rightarrow \infty$. Note that the purity exhibits a minimum of $\varphi=1/3$ at $\gamma=0$, indicating that the domain is neither an acceptor nor a donor. At that value of $\gamma$, the state is given by a homogeneous mixture, represented by $\hat{\rho} = \frac{1}{3} \sum_{\{M\}} {^{M}{\hat{\rho}}}$, for any $q$, where $\{M\}=N,N\pm q$, hence the sum is performed over the three states that generate the Fock space. Higher values of $q$ narrow the curves around zero chemical potential, indicating neutral behavior.

Fig. \ref{fig2:b} reflects the covariance between the domain state $\hat{\rho}$ and $\hat{M}$ as a function of the chemical potential $\gamma$ for different values of $q$. The physical meaning of this magnitude is a measure of the fluctuation-correlation between both operators. The diagram depicts two extreme points: a maximum when $\gamma < 0$ and a minimum when $\gamma > 0$. Also, as $\gamma \rightarrow \pm \infty$, $\hat{\rho}$ and $\hat{M}$ becomes uncorrelated observables, i.e. $\mathrm{Cov}_{\hat{\rho}}(\hat{\rho},\hat{M})$ $\xrightarrow[\gamma \longrightarrow \pm \infty]{}$ $0$. Due to the covariance between $\hat{\rho}$ and $\hat{M}$ is an odd function of $\gamma$, these operators are also uncorrelated at $\gamma=0$. Additionally, as the value of $q$ increases, the curves become narrower, and the correlation between the operators can reach very high values (strong statistical correlation); however, it tends to decrease rapidly as long as $\gamma \rightarrow \pm \infty$.


\section{Concluding remarks}

The quantum state of the molecular domain, represented by the density matrix $\hat{\rho}$, has been expanded to include
three states with different particle numbers, allowing us to describe the electron distribution accurately. The theoretical formulation of quantum molecular properties has been presented using the Quantum Expectation Identity, a purely quantum mechanical theorem. Then, the fluctuation-dissipation identities presented in this work enable us to establish the statistical variance of the particle number, from first principles, as a function of the control parameter $\gamma$. We have focused on three main quantities of the domain: the electronic population, the chemical potential (or control parameter) $\gamma$, and the maximum capacity for accepting or donating charge with the surrounding domains. 
About the quantum state used to describe the molecular domain: starting from $\hat{\rho}$ it is possible, by use of contraction operations, to obtain the $1$-particle reduced density matrix, which would allow us to visualize the collective behavior of the electron density \cite{Coleman_Yukalov_book}.

The purity of the ensemble was described by Eq. (\ref{purity}). Its derivative with respect to the chemical potential allowed us to obtain an analytical expression for the covariance between the density matrix and the number operator as a function of $\gamma$. The importance of the statistical correlation between these observables relies on the fact that once an expression for $\mathrm{Cov}_{\hat{\rho}}(\hat{\rho},\hat{M})$ is known, we can visualize regimes for both strong correlation and for decorrelation. 

The magnitudes discussed in this report have well-defined analytical forms that allow for their behavior to be illustrated graphically.

The next step in our research will involve using this information to compute statistical correlations between different electronic properties and to analyze the fluctuations of charge within a domain.

\section{Acknowledgments}

\noindent R. C. B. has received financial support from Project PIP No. 11220090100061 (Consejo Nacional de Investigaciones
Científicas y Técnicas, Rep\'ublica Argentina). R. C. B. also expresses gratitude to the Department of Physics at the
Facultad de Ciencias Exactas y Naturales, Universidad de Buenos Aires, for providing the necessary facilities.\\

\noindent B.M. sincerely thanks Miss Fernanda Paola for her encouragement and for the time she provided for interesting conversations, around a couple of coffees, about physics, education and related topics.

\section*{References}

\bibliography{references}
\bibliographystyle{aip}

\end{document}